\documentclass[aps,prb,twocolumn,showpacs]{revtex4}

\usepackage{graphicx}% Include figure files
\usepackage{dcolumn}% Align table columns on decimal point
\usepackage{bm}% bold math
\usepackage[normalem]{ulem}

\begin{document}

\title{Second layer of H$_2$ and D$_2$ adsorbed on graphene
}  

\author{M. C. Gordillo}
\affiliation{Departamento de Sistemas F\'{\i}sicos, Qu\'{\i}micos 
y Naturales, Facultad de Ciencias Experimentales, Universidad Pablo de
Olavide, Carretera de Utrera km 1, 41013 Sevilla, Spain}

\author{J. Boronat}
\affiliation{Departament de F\'{\i}sica i Enginyeria Nuclear, 
Universitat Polit\`ecnica de Catalunya, 
Campus Nord B4-B5, 08034 Barcelona, Spain}

\date{\today}

\begin{abstract}
We report diffusion Monte Carlo calculations on the phase diagrams of 
$para$-H$_2$ and $ortho$-D$_2$ adsorbed on top of a first layer 
of the same substances on graphene.  We found that 
the ground state of the second layer is a triangular incommensurate solid
for both isotopes.
%{\bf for both isotopes} 
%the ground state
%{\bf consists of two triangular incommensurate solids of unequal density}. 
%is a quasi two-dimensional incommensurate triangular
%solid {\bf similar to  the one underneath}.
The densities for promotion to a second layer and for the 
onset of a two-dimensional solid  on that second layer compare favorably to 
available experimental data  in both cases.   
\end{abstract}
 
%\keywords{}

\maketitle

\section{INTRODUCTION}

Graphite and graphene are closely related forms of carbon in which the
atoms  are located in the nodes of a honeycomb lattice. The main difference
is that graphite is formed by a whole stack of the two-dimensional carbon
sheets that constitute a single graphene
layer.~\cite{science2004,pnas2005}  Adsorption of gases on top of graphene
and graphite  are expected to show similar trends. This is in fact what has
been shown by computer simulations of the phase diagrams of adsorbed
quantum gases on graphene,  both in the
first,~\cite{prl2009,prb2010,prb2011a,prb2011b,carmen,jltp2013} and
second~\cite{prb2012a} layers. In general, those calculations render phase 
diagrams which are very close to the experimental ones on 
graphite.~\cite{colebook,grey,grey2,frei1,frei2} The only appreciable
difference  is the binding energy of the adsorbate species on top of the
carbon  structure, bigger in the graphite case. Unfortunately, to our
knowledge,  there are no experimental data yet on adsorption of quantum
gases on  graphene to be compared to.    

Most of that work, both from the experimental and theoretical sides, has
been  devoted to the determination of the phase diagram of the first layer 
of quantum gases and solids adsorbed on top of  graphite.~\cite{colebook}
However, there has been some measurements  of the properties of a second
$^4$He sheet on top of an incommensurate helium  layer, directly in contact
with a graphite surface,~\cite{grey,grey2,chan,cro,cro2}  work that can be
compared to the simulations on the same  subject.~\cite{Whitlock,boninx} On
the other hand, the second layer of molecular hydrogen on top of graphite 
has been less explored. Experimentally, the second layer of molecular
hydrogen and deuterium adsorbed both on  graphite and MgO has been studied
using  calorimetric  measurements.~\cite{vilches1,vilches2,vilches3,doble}  
In particular, Ref.
\onlinecite{doble} studies the promotion to a second layer and the  phase
diagram of pure $para$-H$_2$ ($p$-H$_2$) and $ortho$-D$_2$ ($o$-D$_2$) 
second layers on graphite.  One of the main conclusions of that work is that 
the ground
state of both isotopes in the second layer is a quasi two-dimensional 
solid.  Those calorimetric measurements suggest triple points
at T $\sim$ 6 K for H$_2$ and T $\sim$ 11 K for D$_2$. For lower  
temperatures and densities those solids 
seem to coexist with infinitely diluted gases.    
The main goal of our present work is to perform
diffusion Monte Carlo calculations to determine   the phase diagram
of the second layer of H$_2$ adsorbed on a first layer of  H$_2$ on
graphene,  and of the second layer of D$_2$ adsorbed on the first layer of
D$_2$, also on graphene. The results so obtained will be compared to the
experimental ones for the same systems on graphite,~\cite{doble} the only ones
available, in order
to assess the possible differences.  In particular, we try  to 
ascertain if the oblique structure suggested for D$_2$ in 
Refs. \onlinecite{doble}, from neutron diffraction 
experiments, and in Ref. \onlinecite{liu}, from low-energy
electron-diffraction (LEED) measurements,
is more stable than an arrangement  
consisting of two incommensurate triangular solids of different densities.  

In the next Section, we describe the theoretical method used in the
microscopic study of the adsorbed phases. The results obtained for both
$p$-H$_2$ and $o$-D$_2$ on graphene are shown in Sec. III. Finally, the
main conclusions are discussed in Sec. IV.

\section{METHOD}

The diffusion Monte Carlo (DMC) method is a stochastic technique that
allows us to obtain the  ground state of a zero-temperature system of bosons.
Since the species adsorbed ($p$-H$_2$ and $o$-D$_2$) on graphene are both
bosons,  we can obtain through DMC the arrangement of molecules with the
lowest energy for each surface density. In order to reduce the statistical
variance of the many-body problem, the
algorithm uses a guiding wave function $\Phi$ 
which  enhances the occupation probability in places where the hydrogen
density is expected to be large.~\cite{boro94} In general,  $\Phi$ depends on the
coordinates of all atoms or molecules in  the simulation cell. However,
in this work, we will consider that the carbon atoms  of the 
graphene layer are kept in fixed positions, what means that their overall
effect  on the hydrogen molecules can be described as an external
potential. Within this approximation,  the guiding wave function depends
only on the positions of the $N$ hydrogen molecules   ({\bf
r}$_1$,$\ldots$,{\bf r}$_N$). We chose, 
\begin{eqnarray}
\Phi({\bf r}_1,\ldots,{\bf r}_N) & = & \Phi_J({\bf r}_1,\ldots,{\bf r}_N)
\Phi_1({\bf r}_1,\ldots,{\bf r}_{N_1})  \nonumber \\
& & \times \ \Phi_2({\bf r}_{N_1+1},\ldots,{\bf r}_N) \ ,
\label{phitot}
\end{eqnarray}
where $N_1$ ($N_2=N-N_1$) is the number of molecules adsorbed in the first
(second) layer, and $\Phi_J$ 
is a Jastrow wave function used to take into account the H$_2$-H$_2$ and D$_2$-D$_2$ 
correlations.~\cite{prb2010,carmen} In particular,  
\begin{equation}
\Phi_J({\bf r}_1,\ldots,{\bf r}_N) = \prod_{1=i<j}^{N} \exp \left[-\frac{1}{2} 
\left(\frac{b}{r_{ij}} \right)^5 \right] \ ,
\label{sverlet}
\end{equation}
$r_{ij}$ being the distance between the center of mass of hydrogen
molecules  (considered  as spheres and interacting through the isotropic 
Silvera and Goldman potential~\cite{silvera}).  This potential was
developed  to describe the hydrogen-hydrogen interaction in bulk and it
was successfully used to reproduce the phase diagram of the first layer of both
H$_2$ and D$_2$ on graphene.~\cite{prb2010,carmen} The variational
parameter $b$ in Eq. (\ref{sverlet})   was taken to be $3.195$ \AA$ $ for
both H$_2$ and D$_2$, in agreement with  variational optimizations done in
previous simulations  of similar systems.~\cite{prb2010,carmen}  

The purpose of the remaining terms in Eq. (\ref{phitot}), $\Phi_1$ and $\Phi_2$, is 
to describe the localization of hydrogen molecules in the first and second
layers on top of graphene, respectively, and also the correlations due to
the carbon-hydrogen interactions. Concretely, 
\begin{eqnarray}
\Phi_n({\bf r}_1, {\bf r}_2, \ldots, {\bf r}_{N_n})  =  
\prod_i^{N_n}  \prod_J^{N_C} \exp \left[ -\frac{1}{2} \left( \frac{b_{{\text
C}}}{r_{iJ}} \right)^5 \right] \\
\prod_i^{N_n} \exp (-a_n (z_i-z_n)^2) \nonumber \ 
\label{phicapa}
\end{eqnarray} 
with $n =1$, 2 for the first and second hydrogen layer, respectively, and
$N_n$ the number of  molecules in each layer. $r_{iJ}$ represents the
distance between the center or mass of each molecule, $i$, and each of the
$N_C$ carbon atoms, $J$,  in the graphene layer. Each hydrogen  molecule
interacts with each of those carbon atoms by a potential of 
Lennard-Jones type, whose  parameters are taken from Ref. \onlinecite{coleh2}.
This part of the guiding wave function is again similar  to the one used in
previous works to describe the first layer of H$_2$ (Ref.
\onlinecite{prb2010}) and D$_2$ (Ref. \onlinecite{carmen}) on graphene, to
the point that we took the same $a_1$ and $z_1$ parameters as in those
works. Thus, $a_1 = 3.06$ \AA$^{-2}$ for H$_2$ and a$_1 = 5.2$  \AA$^{-2}$
for D$_2$.  $b_C= 2.3$ \AA and $z_1 = 2.9$ \AA for both hydrogen isotopes.
For hydrogen molecules on the second layer,  $b_C$ was kept constant, and
$a_2$ and $z_2$ were variationally optimized. The optimal values were 
$a_2 = 1.53$
\AA$^{-2}$  for both H$_2$ and D$_2$, and $z_2 = 6$ \AA (H$_2$) and $5.8$ \AA
(D$_2$). 

When the phase to be described is a quasi two-dimensional solid, 
$\Phi_n$ is multiplied by a Nosanow term, 
\begin{equation}
\prod_i \exp \{ -c_n [ (x_i-x_{\text{site}})^2+ (y_i - y_{\text{site}})^2 ]
\} \ ,
\label{trialsol}
\end{equation}
where ($x_{\text{site}},y_{\text{site}}$) are the crystallographic
positions of the solid lattice. The $c_n$ parameters were taken to be
the same for $n$= 1,2, i.e., a linear interpolation between the values
corresponding to densities in the range 0.08 \AA$^{-2}$ ($c_n=  0.61$
\AA$^{-2}$) and 0.10 \AA$^{-2}$ ($c_n =  1.38$ \AA$^{-2}$)    for 
H$_2$,~\cite{prb2010} and between 0.08 \AA$^{-2}$ ($c_n= 1.11$ \AA$^{-2}$) and
0.11 \AA$^{-2}$ ($c_n= 2.93$ \AA$^{-2}$)  in the case of 
D$_2$.~\cite{carmen} If the hydrogen density within the considered layer was not
in those ranges, we used the linear extrapolated $c_n$ value.   

To model the second layer of molecular hydrogen on top of a first layer of the
same substance, we followed closely the prescription  of Ref.
\onlinecite{prb2012a}, in which a second layer of $^4$He on graphene was
simulated. Basically, for a fixed total hydrogen density, we considered
only the arrangement for which the total energy per molecule was lower. In
practice, this means that  we have to take a fixed solid density for the
first layer, and to change the number of molecules (if we have a liquid) or the
lattice constant (if we have a solid) in the second layer.
The structure of the second layer solid was assumed to be incommensurate
with respect to the one on the first layer, i.e., the phase diagram was assumed
to be of the same type as that of helium on graphene. However, to verify that 
this was so, we considered also the oblique commensurate structure proposed 
for D$_2$ from neutron diffraction \cite{doble} and LEED \cite{liu}. We checked the 
stability of that structure for both isotopes.   
To be able to treat
incommensurate second layers, we used different periodic  boundary
conditions for the first and second layers. No exchange of molecules
between the first and second layer was allowed. Importantly, we did not
fix the positions of the molecules closest to the graphene surface, i.e.,
we took into account the zero point motion  of all the hydrogen molecules. For both
isotopes, we considered hydrogen densities up to those of 
promotion to a third  layer, obtained experimentally for the same systems on
graphite.~\cite{doble}

\section{RESULTS}

\subsection{H$_2$}

\begin{figure}[b]
\begin{center}
%\vspace*{0.5cm}
\includegraphics[width=7.5cm]{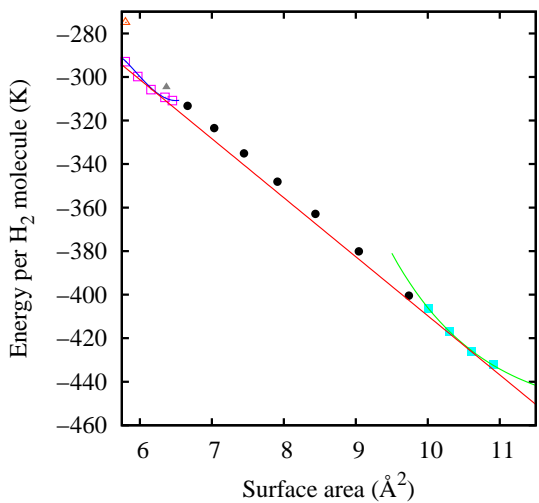}
\caption{(Color online)
Maxwell construction (straight line) to determine the limits of the phase
diagram of a second layer of H$_2$ on top of the same substrate on
graphene.  Full squares, energy per H$_2$ molecule on a first layer
triangular incommensurate solid; full circles, same data for  a 
second layer liquid on a triangular solid; open squares, a triangular
incommensurate solid on top  of first layer solid. The open triangle
represents the energy per H$_2$ molecule of the oblique structure proposed in
Ref. \onlinecite{doble}, while the full triangle is the energy of a second
layer 4/7 structure. The curves are three-order polynomial fits to the
simulation results.  
}
\label{energy1}
\end{center}
\end{figure}

The phase diagram of the second layer of H$_2$ on graphene can be extracted
from the data displayed in Fig.~\ref{energy1}.    The full squares
correspond to the energy per molecule of a triangular incommensurate solid
adsorbed on a single layer  on top of graphene, and were taken from Ref.
\onlinecite{prb2010}. When we considered only a hydrogen layer,
incommensurate  means that there is no registry of the hydrogen molecules
with respect to the carbon sheet. The third-order polynomial fit displayed
on top of them is  a guide-to-the eye. To study second layer structures, we
put on top of one of these incommensurate structures a set of hydrogen
molecules  described by a guiding wave  function with $c_2 = 0$
(\ref{trialsol}). The dimensions of
the simulation cells were determined by the density of  incommensurate
solid in the first layer since, in all cases, the simulation cells
comprised 120 H$_2$ molecules close to the  graphene surface.  Then, we put
enough hydrogen on top of them to produce surface areas (the inverse of
density)  in the range displayed in Fig.~\ref{energy1}. To be sure that
the energy per molecule considering both hydrogen sheets was the minimum
one,  we performed several sets of simulations with different
incommensurate solid densities (in the first layer). 
In particular, we considered  $0.090$,
$0.095$, $0.10$ and $0.105$ \AA$^{-2}$. Our results indicate that  on the second layer, the arrangement with
lowest energy per H$_2$ molecule is a liquid on top of a solid whose
density is $0.095$ \AA$^{-2}$. Those results are displayed in
Fig.~\ref{energy1} as full circles.  The lowest limit of the surface area
corresponds approximately to the experimental value for H$_2$ promotion to
a third layer ($5.80$ \AA$^2$).~\cite{doble}

We modeled the second layer incommensurate triangular solid on the same
principles, i.e., we considered the same densities as for the liquid
case  in the first hydrogen sheet, and distributed the atoms on the second
layer as to produce the total densities displayed as surface areas   in
Fig.~\ref{energy1}. Here, incommensurate means that the second layer is
registered neither with respect to the first layer nor with the
underlying graphene. The minimum energies per H$_2$ molecule corresponded to an
arrangement in which the lower layer density was  $0.10$ \AA$^{-2}$; 
those data are displayed in Fig.~\ref{energy1} as open squares. The line
on top of them represents also a third-order polynomial fit.  
We considered also two second layer commensurate (with
respect to the first layer) solids: a
$4/7$ lattice,~\cite{grey2} and an oblique bilayer  structure,~\cite{carneiro} 
both proposed originally to describe second layers of
$^4$He on graphite. The latter was suggested to be  stable for the second
layer of H$_2$~\cite{doble} and D$_2$ ~\cite{doble,liu} on graphite. The energy per H$_2$ molecule
of those registered phases is represented by a full  and open 
triangle, respectively.
We can see that both structures are metastable with respect to a set of two
incommensurate layers, since their  energies per molecule are larger. 
Their corresponding energies per H$_2$ molecule are -274.9 $\pm$ 0.1 K (oblique structure),
and -304.6 $\pm$ 0.1 K ($4/7$), versus -293.0 $\pm$ 0.1 K and -309.38 $\pm$ 0.08 K of the
corresponding incommensurate structures of the same densities (0.173 and 0.157 \AA$^{-2}$, respectively). 
A $7/12$ commensurate solid, also proposed to be stable for helium,~\cite{boninx} 
and not displayed for simplicity, was also  considered and found to be
unstable with respect to a second layer incommensurate solid.   

With all of that in mind, we can draw a double-tangent Maxwell construction
to obtain the phase diagram of the second layer of H$_2$ on graphene.  The
slope of that line is minus the internal pressure at which the transition
takes place, and has to be positive for stable arrangements. In addition,
if several transitions are possible, one has to consider only the corresponding to the lowest
pressure. The straight line  displayed in     
Fig.~\ref{energy1}, that joints a single layer incommensurate solid and
its second layer counterpart fulfills all the necessary requirements.  We
can see that the energy per H$_2$ molecule of a second layer liquid (full
circles) is always above the double-tangent line. This means that  a
second layer liquid is unstable with respect to a mixture of a single layer
solid and another with two incommensurate sheets. The  surface areas at
which the slopes of both three-order polynomial fits are the same define
the stability limits for the phases involved in the transition. In our
case, those correspond to an upper density limit for a single layer solid
(equivalent to a second layer promotion)  of $0.094 \pm 0.002$ \AA$^{-2}$,
and a lowest density value of $0.160 \pm 0.002$ \AA$^{-2}$ for a two
layered solid. Both results are in excellent agreement with the
 calorimetric values of H$_2$ on graphite:~\cite{doble} $0.094$ 
and $0.165$ \AA$^{-2}$.
The chemical potential for H$_2$ at the monolayer-bilayer transition 
derived from our simulations was -161 $\pm$ 2 K. This means that 
a second layer of H$_2$ is still stable with respect to the formation of
a bulk H$_2$ crystal, whose ground state chemical potential is -92.3 K \cite{ost}.
This also means 
that the second layer solidifies at H$_2$ densities
as low as $0.060$ \AA$^{-2}$. As in $^4$He,~\cite{grey,grey2,prb2012a} 
the first layer suffers a compression that
produces a $\sim 6$\% increase of its density 
upon the adsorption of the second hydrogen sheet. 

\begin{figure}[b]
\begin{center}
%\vspace*{0.5cm}
\includegraphics[width=7.5cm]{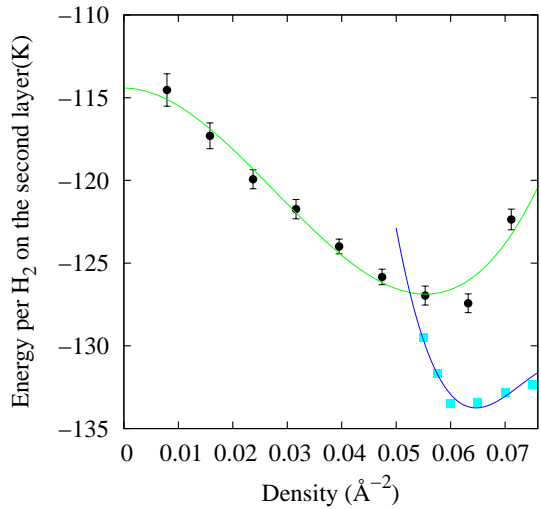}
\caption{(Color online) Energy per H$_2$ for molecules located
on the second layer. Full circles are the results of a liquid phase, and
full squares, for a triangular incommensurate solid one. The error bars
are similar in both cases and for simplicity they are only shown in the
liquid case. Lines are three-order polynomial fits to the respective data. 
}
\label{energy2}
\end{center}
\end{figure}

We also compared our simulation results for H$_2$ on a second layer to
similar data obtained for a pure two dimensional (2D) system.   In
Fig.~\ref{energy2}, we show the energy per H$_2$ molecule as a function of
density, but only for the molecules on that second sheet.   This energies
are taken from the same set of simulation results as the ones displayed
in Fig.~\ref{energy1}, i.e., the molecules of the first layer are not
kept frozen.  A  glance to the previous figure indicates that the
binding energies of these molecules are smaller than the ones 
located in the first layer by at least a factor of two. This is true for 
both liquid
(full circles) and triangular incommensurate (full squares)
phases.    The lines on top of each set of symbols are least-squares
fits to the expression 
\begin{equation} 
E/N  = (E/N)_0 + a (\rho- \rho_0)^2 + b (\rho - \rho_0)^3 \ , 
\label{fit}
\end{equation}
where $\rho$ is the hydrogen density in that second layer and $\rho_0$
stands for the density at which the energy per H$_2$ molecule has a minimum
($(E/N)_0$).  The parameters obtained for the liquid and solid phases, together
with their first layer (taken from Ref. \onlinecite{prb2010}) and pure 
two-dimensional (from Ref. \onlinecite{claudio}) counterparts, are shown in 
Table \ref{table1}. Those results indicate that a pure two-dimensional
system of H$_2$ molecules is a worse description for the second
than  for the first layer of H$_2$ on graphene, and that 
the 2D equation of state is
worse for a liquid phase than for a solid one.   For instance, the
energy differences, once subtracted the infinite-dilution energy
($E_{\infty}$),
between a 2D system and a first layer of H$_2$ on graphene  are $\sim 6.5$
\% for a liquid, and $\sim 5$ \% for a solid. In comparison, those same
differences between a 2D and a second layer system are $\sim 71$ and 22 \%,
respectively. The same, but to a lesser extend, can be said of the
differences between the values of $\rho_0$: $\sim 2.3$ \% for the 2D and
first layer solids difference, gap that increases up to $\sim 4.2$ \% if
instead of a first layer we have a second one.  Those same percentages
grow to $\sim 6.5$ \% and 15 \% for the same comparisons for liquid
phases. Moreover, Table \ref{table1} gives us another interesting
piece of information: the upper solid is more stable with respect to a
liquid arrangement than in a flat structure or a first layer sheet.
This is because the the energy difference at the respective
equilibrium densities is  highest in the second layer case. The corresponding values are 6.8 K
(second layer), versus 2.023  K (2D), and 1.8 K (first layer). We can also see that the
density minimum for the second-layer solid phase ($0.0646 \pm 0.0009$
\AA$^{-2}$) is comparable to the equilibrium density
mentioned above ($0.060 \pm   0.002$ \AA$^{-2}$), obtained from the Maxwell
construction including the whole system.

\begin{table}
\caption{Density and energy per molecule for the phases displayed in
Fig.~\ref{energy2}    from fits using  Eq. (\ref{fit}). The
results for a pure two dimensional system (2D, Ref. \onlinecite{claudio})
and a first layer of H$_2$ on top of graphene (gr, Ref.
\onlinecite{prb2010}) are also shown for comparison.  $E_{\infty}$ is the energy
per H$_2$ molecule in the infinite dilution limit for each system, $(E/N)_0$ is the
minimum energy per H$_2$ molecule and $\rho_0$ is the density that corresponds to
that energy.  }
\begin{tabular}{ccc} \hline
           & Liquid & Solid  \\ \hline
           &  2D    &       \\ \hline
$E_{\infty}$    (K) & 0   &  \\
$(E/N)_0$    (K) & -21.43 $\pm$ 0.02 & -23.453 $\pm$ 0.003 \\
$(E/N)_0-E_{\infty}$    (K) & -21.43 $\pm$ 0.02 & -23.453 $\pm$ 0.003 \\
$\rho_0$ (\AA$^{-2}$)  & 0.0633 $\pm$ 0.0003 &  0.0673 $\pm$ 0.0002 \\ \hline
           &  gr           \\ \hline
$E_{\infty}$  (K) &  -431.79 $\pm$  0.06  &    \\
$(E/N)_0$  (K) &  -451.88 $\pm$  0.03  &  -454.1 $\pm$ 0.3  \\
$(E/N)_0-E_{\infty}$  (K) &  -20.09 $\pm$  0.07  &  -22.3 $\pm$ 0.3 \\
$\rho_0$ (\AA$^{-2}$)  & 0.05948 $\pm$ 0.00005 &  0.0689 $\pm$ 0.0006  \\ \hline
           &  second layer    &       \\ \hline
$E_{\infty}$  (K) &  -114.4 $\pm$  0.6  &    \\
$(E/N)_0$  (K) &  -126.9 $\pm$  0.5  &  -133.7 $\pm$ 0.3  \\
$(E/N)_0-E_{\infty}$  (K) &  -12.5 $\pm$  0.8  &  -19.3 $\pm$ 0.7 \\
$\rho_0$ (\AA$^{-2}$)  & 0.055 $\pm$ 0.001 &  0.0646 $\pm$ 0.0009 \\
 \hline
\end{tabular}
\label{table1}
\end{table}

\subsection{D$_2$}

\begin{figure}[b]
\begin{center}
%\vspace*{0.5cm}
\includegraphics[width=7.5cm]{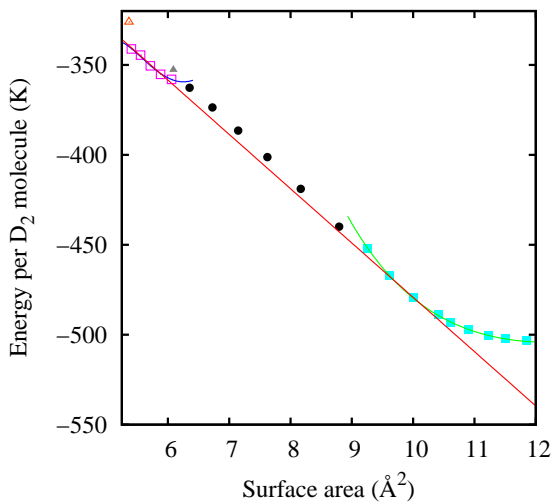}
\caption{(Color online) Same as in Fig.~\ref{energy1}, but for D$_2$
instead of H$_2$. The lower limit for the surface density is fixed as the
inverse of the experimental density for a promotion of D$_2$ to a third
layer ($5.8$ \AA$^{2}$, Ref. \onlinecite{doble}).  
}
\label{energy3}
\end{center}
\end{figure}

This subsection will closely mirror the previous one since we studied the
second layer of D$_2$ on graphene following the same steps. 
Our results are summarized in
Fig.~\ref{energy3}, where all the symbols and lines have a similar meaning
to those of Fig.~\ref{energy1}. The only difference is that the
structures with the minimum  energy per deuterium molecule are those whose
first layer density is $0.105$ \AA$^{-2}$ for both  the second layer solid and
liquid phases. To be sure of that, densities in the range $0.095$ to
$0.110$ \AA$^{-2}$ were tested.  The first layer results (full squares) are
now taken from Ref. \onlinecite{carmen}, and the triangles represent the
same registered phases  suggested above for H$_2$. As one can see, these
commensurate phases are still unstable
with respect to a set of two incommensurate solid deuterium layers.       
As in H$_2$, we have also that the energies per D$_2$ molecule for a second layer
liquid are above the double-tangent Maxwell construction. This means that
the phases in equilibrium are again a first layer incommensurate triangular
solid of density $0.100 \pm 0.002$ \AA$^{-2}$ and a second layer 
incommensurate solid whose total density is $0.175 \pm 0.002$ \AA$^{-2}$.
Both results are again in excellent  agreement with the  calorimetric data
of Ref. \onlinecite{doble} on graphite: $0.099$ \AA$^{-2}$ for second layer promotion,
and $0.178$ \AA$^{-2}$ for the minimum density at which the double solid is
stable. Since in this last case, the density of the lower layer is  $0.105$
\AA$^{-2}$, we can state that  there is also a compression of the first layer
of around a 5 \% upon the adsorption of a second layer of D$_2$ on top of
D$_2$.  As in the case of H$_2$, our simulation results allow us to calculate
the chemical potential of D$_2$ at the monolayer-bilayer transition. The result
was -180 $\pm$ 2 K, a healthy 30\% larger than the corresponding to bulk D$_2$.
We found also that both the oblique commensurate structure and the $4/7$ one were
unstable with respect to the double incommensurate arrangement, as can be 
seen in Fig.~(\ref{energy3}). The values of their energies per 
D$_2$ molecule were -326.1 $\pm$ 0.1 K and -352.7 $\pm$ 0.1 K, versus 
-341.1 $\pm$ 0.1 K and -358.0 $\pm$ 0.1 K for the double incommensurate
for the same densities (0.186 and 0.164 \AA$^{-2}$, respectively).

\begin{figure}[b]
\begin{center}
%\vspace*{0.5cm}
\includegraphics[width=7.5cm]{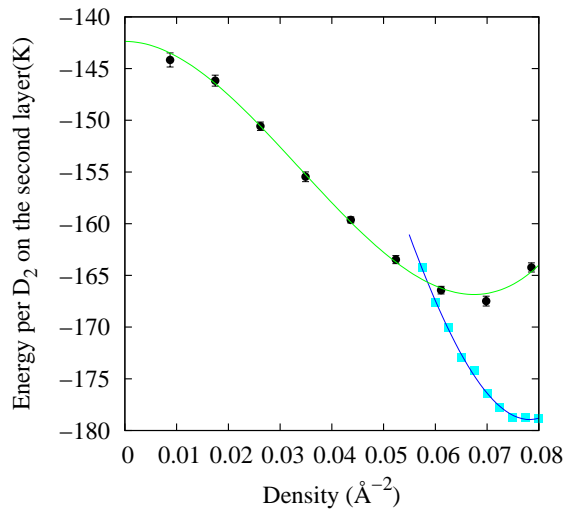}
\caption{(Color online) Same as in Fig.~\ref{energy2}, but for D$_2$ 
instead of H$_2$.
}
\label{energy4}
\end{center}
\end{figure}

We can also study the second layer by itself and compare the results to
those of a first layer of D$_2$ adsorbed on graphene, and to a pure 2D  system. 
That
can be done with the help of Fig.~\ref{energy4} and Table \ref{table2}.
The conclusions we can draw from this set of information are similar to
those already described in the H$_2$ case: the equilibrium densities
are essentially compatible with each other, even more so in the solid case,
and the energy per  D$_2$ molecule differences between liquid and a solid phases are
larger in the case of a second layer solid than for a single deuterium
sheet. The  equilibrium density of the second layer solid is also comparable to 
the
one extracted from the  Maxwell construction for the entire system
($0.078$ versus $0.070$ \AA$^{-2}$). 
    
\begin{table}
\caption{Density and energy per D$_2$ molecule for the phases displayed in
Fig.~\ref{energy4}, obtained by the same means as those of H$_2$. The results 
for a pure two-dimensional system (2D, Ref.
\onlinecite{claudio}, with only data for the solid phase) and a first layer
of D$_2$ on top of graphene (gr, Ref. \onlinecite{carmen} for the liquid
and this work of the solid) are also shown for comparison.  The variables
have the same meaning as  in Table \ref{table1}. 
}
\begin{tabular}{ccc} \hline
           & Liquid & Solid  \\ \hline
           &  2D    &       \\ \hline
$E_{\infty}$    (K) & 0   &  \\
$(E/N)_0$    (K) & --- & -42.305 $\pm$ 0.005 \\
$(E/N)_0-E_{\infty}$    (K) & --- & -42.305 $\pm$ 0.005 \\
$\rho_0$ (\AA$^{-2}$)  & --- &  0.0785 $\pm$ 0.0002 \\ \hline
           &  gr           \\ \hline
$E_{\infty}$  (K) &  -464.87 $\pm$  0.06  &    \\
$(E/N)_0$  (K) &  -497.2 $\pm$  0.9  &  -504.2 $\pm$ 0.08  \\
$(E/N)_0-E_{\infty}$  (K) &  -32.3 $\pm$  0.9  &  -39.3 $\pm$ 0.1 \\
$\rho_0$ (\AA$^{-2}$)  & 0.064 $\pm$ 0.001 &  0.0799 $\pm$ 0.0002  \\ \hline
           &  second layer    &       \\ \hline
$E_{\infty}$  (K) &  -142.4 $\pm$  0.5  &    \\
$(E/N)_0$  (K) &  -166.8 $\pm$  0.5  &  -178.9 $\pm$ 0.6  \\
$(E/N)_0-E_{\infty}$  (K) &  -24.4 $\pm$  0.7  &  -36.5 $\pm$ 0.8 \\
$\rho_0$ (\AA$^{-2}$)  & 0.068 $\pm$ 0.002 &  0.078 $\pm$ 0.003 \\
 \hline
\end{tabular}
\label{table2}
\end{table}

\section{CONCLUSIONS}

In this work, we have studied the complete phase diagram of the second
layer of both H$_2$ and D$_2$ adsorbed on  top of a single graphene layer.
Since we used a set diffusion Monte Carlo calculations, the results
correspond to the zero-temperature ground state of the system. To obtain
the real  stability limits,  we had to perform Maxwell constructions
between phases that comprised  one and two hydrogen layers, but we found
that if we used only the data corresponding to a second  one, the
description is good enough to reproduce the solid equilibrium of the
complete two-sheet system. Our results for the promotion density to the
second layer and the minimum density in this second layer agree
satisfactorily with available  calorimetric results for graphite,
pointing to the accuracy of both the method used and the interaction
potentials entering in the Hamiltonian.  That our results on graphene are
comparable to the experimental ones on graphite also means that both
surfaces are basically equivalent as absorbents, being the only possible
difference the binding energy of the hydrogen molecules to the carbon
surface. 

 However, calorimetric measurements only give the total
density for the onset of  a solid structure at T $\rightarrow$ 0. From Ref.
\onlinecite{doble}, those densities  appear to be 0.165 (H$_2$) and 0.178
(D$_2$) \AA$^{-2}$. Both are lower than the ones assigned in the same work
to the oblique structures: 0.173 (H$_2$) and 0.186 (D$_2$) \AA$^{-2}$,
respectively.  The fact that the former densities are compatible with our
simulation results supports our suggestion of a double incommensurate solid
as the structure for the inferred solid phase. Moreover, a comparison of
the energies per molecule for the oblique solid and the two sets of
triangular layers at the same densities indicates that, at least for $T =0$,
the  commensurate arrangement is not stable. One could speculate that the
disagreement between calorimetric data, in one side, and LEED and neutron
diffraction, on the other, could be originated by the tiny difference between
the diffraction patterns of a double
incommensurate structure and a double oblique layer.~\cite{doble,liu}  
In fact, the oblique phase was proposed in the past as the ground state of
a second layer of $^4$He on graphite in neutron diffraction 
studies,~\cite{carneiro} but further calorimetric measurements concluded
that triangular structures were preferred. This could be also the case for 
hydrogen. On the other hand, our results rely on empirical potentials that  
have been  used to reproduce
reasonably well the equation of state of the first layer of both H$_2$ and D$_2$ 
on graphene,~\cite{prb2010,carmen} but we can not exclude that, in the
future,  more
elaborate interactions could change our predictions for the second layers.

On a different note, from the behavior of this second layer,  we can say that 
when the
number of adsorbed layers grows, the entire arrangement  becomes more
``solid-like" than ``liquid-like".      
The reason is that, even though the binding energy in the second layer is
lower than in the first one, the incommensurate solid structure is much
more stabilized with respect to a liquid than  in the first layer case.
This is probably the reason why the experimental 
 critical points for the liquid-vapor coexistence
regions of 
H$_2$ and D$_2$ adsorbed on graphite  approach those of the bulk 
solids as the number of adsorbed layers grows,
~\cite{doble} making impossible to obtain a stable liquid
by reducing the dimensionality of the system.

\acknowledgments
We acknowledge partial financial support from the
Junta de Andaluc\'{\i}a group PAI-205 and Grant FQM-5985, DGI (Spain)
Grants FIS2010-18356 and FIS2011-25275, and Generalitat de Catalunya 
Grant 2009SGR-1003.

\end{document}